# DENOISE IN THE PSEUDOPOLAR GRID FOURIER SPACE USING EXACT INVERSE PAEUDOPOLAR FOURIER TRANSFORM

*Fan Jun Wei*

*Institute of Physics, Academia Sinica, Taipei, Taiwan*

**ABSTRACT**

In this paper I show a matrix method to calculate the exact inverse pseudopolar grid Fourier transform, and use this transform to do noise removals in the *k* space of pseudopolar grids. I apply the Gaussian filter to this pseudopolar grid and find the advantages of the noise removals are very excellent by using pseudopolar grid, and finally I show the Cartesian grid denoise for comparisons. The results present the noise to signal ratio and the variance are better when doing noise removals in the pseudopolar grid than the Cartesian grid. The noise removals of pseudopolar grid or Cartesian grid are both in the *k* space, and all these noises are added in the real space. The added noises are white noises.

*Index Terms*—exact inverse, pseudopolar, Fourier transform, denoise

## 1. INTRODUCTION

Coherent X ray and electron microscopic are usually used to capture 3D image, virus or nano particles [1] are most common materials. The atomic resolution is the new goal of this kind of 3D images. The 3D image is constructed by a 2D image set, and these 2D images are obtained by experimental data in equal angels usually. One important thing of the signal progressing of this 2D imaging is affected by noise very sensitively. Besides the signals are not continuous states but discretized states. People did noise removals in the *k* space for a long time and very successful by using a filter of Cartesian grid. When the 3D imaging are more and more important, the Cartesian coordinate is impossible to process 3D imaging in the discretized grids [5]. A coordinate named pseudopolar grid suggested by Prof. Miao etc. is in equal slopes in the *k* space to solve 3D image problem in recent years [2]. This type of grid has been explored by many since the 1970s. Although we can do the forward pseudopolar grid Fourier transform, but the form of exactly inverse pseudopolar grid Fourier transform is hard to be found and not easy to be used [3-4].

## 2. THE PSEUDOPOLAR GRID IN THE *k* SPACE

The pseudopolar Fourier transform is based on a definition of a polar like 2D grid that enables fast Fourier computation. An inverse transform can be defined and implemented by solving a corresponding least square problem in conventional way. Eq. (1) shows the form of the least square equation for the conventional way. Let *u* be the 2D variables need to be solved in real space. *Fu* denotes the pseudopolar Fourier transform of *u*. *v* denotes the known data of two dimensions in the pseudopolar *k* space. Fig. 1 shows the data transformed between real space in Cartesian grids and *k* space in pseudopolar grids both of discretized points. In the pseudopolar grid *k* space, the points are more concentrated in the center and the points are looser outside, these conditions help us to develop useful functions to remove noises.

$$F^{-1} \equiv arg\ \min_u \|Fu - v\|_{2D} \quad (1)$$

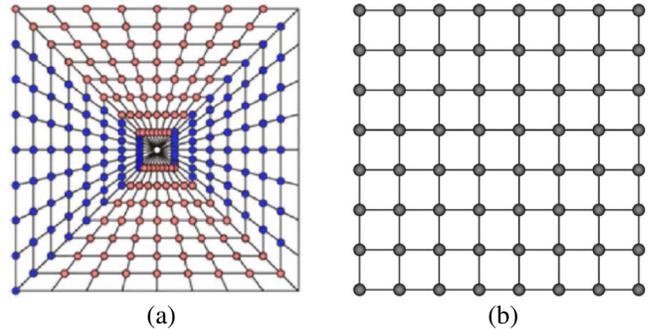

Fig 1. The geometrical relationship between a pseudopolar grids in the *k* space (a) and a Cartesian grids in the real space (b). For an *N*×*N* Cartesian grids in the real space, the corresponding pseudopolar grids is defined by a set of 2*N* lines, each line consisting of 2*N* grid points mapped out on *N* concentric squares with *N*=8 in this example. The 2*N* lines are subdivided into a horizontal group (*BH* in blue) defined by $k_y$=slope×$k_x$, where |slope|≤ 1, and a vertical group (*BV* in red) defined by $k_x$=slope×$k_y$, where |slope|≤ 1. The horizontal and vertical groups are symmetric under the interchange of $k_x$ and $k_y$, and Δslope=2/*N*.

## 3. THE METHOD OF EXACT INVERSE PSEUDOPOLAR GRID FOURIER TRANSFORM

The pseudopolar Fourier transform is different from the standard fast Fourier transform. If the forward pseudopolar Fourier transform, the inverse pseudopolar Fourier transform, and the adjoint pseudopolar Fourier transform are denoted $F$, $F^{-1}$ and $F^A$, respectively, throughout his paper, then I can find $F^{-1}F=FF^{-1}=I$, and $F^A F \neq FF^A \neq I$, that means the adjoint transform $F^A$ itself is not orthogonal, and $F^A$ is not equivalent with the inverse transform of $F^{-1}$. The optimization method is the most common way that people do the inverse pseudopolar Fourier transform conventionally, but the accuracy is low as its problem. In this paper I develop a matrix method to do inverse pseudopolar Fourier transform by taking the adjoint pseudopolar Fourier transform $F^A$ as the bridge to find the exact solution. Although the inverse pseudopolar Fourier transform $F^{-1}$ is proved their existence in some references, but there is no reality methods published. My method is smart by taking forward and adjoint pseudopolar Fourier transform mixed together and it's an exact presentation of the inverse pseudopolar Fourier transform of matrix form without any singularities, so there is no more accuracy problem, and the calculation time is also less than the conventional, optimization method. I also can prove the matrix $\left| (BH^A \ BV^A)_{N^2 \times 4N^2} \binom{BH}{BV}_{4N^2 \times N^2} \right|$ has full ranks, so it can be inversed exactly, there are no singularity points. Here is the definition of my inverse pseudopolar Fourier transform matrix in Eq. (3), and the origin definition is shown in Eq. (2).

$$F_{BH}(k_x, s) = \sum_{x=1}^{N} \sum_{y=1}^{N} e^{-\frac{i\pi}{N}[(k_x-N)x + \frac{s}{N}(k_x-N)y]} \quad (2\text{-}1)$$

$$F_{BV}(k_y, s) = \sum_{x=1}^{N} \sum_{y=1}^{N} e^{-\frac{i\pi}{N}[\frac{s}{N}(k_y-N)x + (k_y-N)y]} \quad (2\text{-}2)$$

$$F_{BH}^A(x, y) = \frac{1}{4N^2} \sum_{k_x=1}^{2N} \sum_{s=-N+2, -N+4, \ldots}^{N} e^{\frac{i\pi}{N}[(k_x-N)x + \frac{s}{N}(k_x-N)y]} \quad (2\text{-}3)$$

$$F_{BV}^A(x, y) = \frac{1}{4N^2} \sum_{k_y=1}^{2N} \sum_{s=-N+2, -N+4, \ldots}^{N} e^{\frac{i\pi}{N}[\frac{s}{N}(k_y-N)x + (k_y-N)y]} \quad (2\text{-}4)$$

$$F^{-1} = \left| (BH^A \ BV^A)_{N^2 \times 4N^2} \binom{BH}{BV}_{4N^2 \times N^2} \right|^{-1} (BH^A \ BV^A)_{N^2 \times 4N^2} \quad (3)$$

$$BH_{(k_x,s)\ (x,y)} = e^{-\frac{i\pi}{N}[(k_x-N)x + \frac{s}{N}(k_x-N)y]}$$
$$BV_{(k_y,s)\ (x,y)} = e^{-\frac{i\pi}{N}[\frac{s}{N}(k_y-N)x + (k_y-N)y]}$$
$$BH_{(k_x,s)\ (x,y)}^A = \frac{1}{4N^2} e^{\frac{i\pi}{N}[(k_x-N)x + \frac{s}{N}(k_x-N)y]}$$
$$BV_{(k_y,s)\ (x,y)}^A = \frac{1}{4N^2} e^{\frac{i\pi}{N}[\frac{s}{N}(k_y-N)x + (k_y-N)y]}$$

where the $BH$, $BV$, $BH^A$, and $BV^A$ in Eq. (3) are the elements of $F^{-1}$ and A means adjoint [6].

## 4. DENOISE IN THE PSEUDOPOLAR GRID FOURIER SPACE

First I choose the picture of Lena's eye as my example picture, has $N \times N = 54 \times 54$ pixels. Second I add white noise in this picture in the real space, and the noise/signal=0.0723.

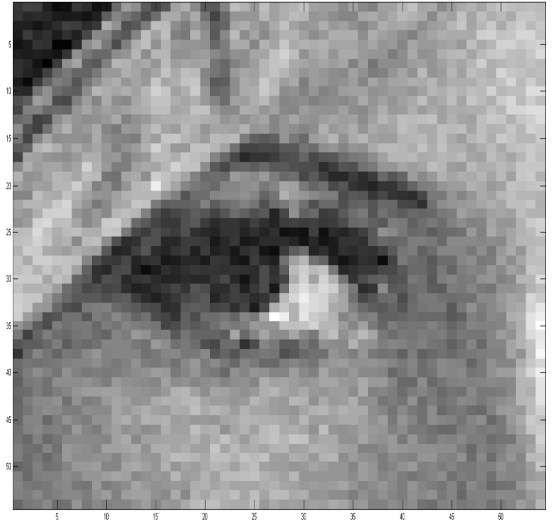

Fig 2. The picture of Lena's eye, that has $N \times N = 54 \times 54$ pixels, is added white noise in the real space, and the noise/signal=0.0723.

Then I do forward pseudopolar Fourier transform of Fig 2. Then I apply a Gaussian filter to this transformed picture in the $k$ space. Now I then do my exact inverse matrix method to get the real space picture of Lena's eye. I calculate the noise/signal ratio by using this formula Eq. (4), and of course the compared pictures in the real space are all normalized.

$$\frac{noise}{signal} \equiv \frac{\sum_{x,y} ||f(x,y)| - |F(x,y)||}{\sum_{x,y} |F(x,y)|} \quad (4)$$

where $F(x,y)$ represents the image of the original Lena's eye without added any noise in the real space and $f(x,y)$ also represents the image in the real space but this image first after is treated a Gaussian filter in the pseudopolar Fourier space and then be calculated by my exact inverse matrix method. I find the noise/signal=0.0525. The Gaussian filter is usually a circle in the $k$ space of a Cartesian grid, but due to the property of pseudopolar grid, I can intuitively take the square filter to replace the circle filter in the pseudopolar grid. This type filter help us more efficient design the function of the filter and the parameters of the square filter that I choose in the pseudopolar $k$ space makes the lowest noise/signal of noise removals.

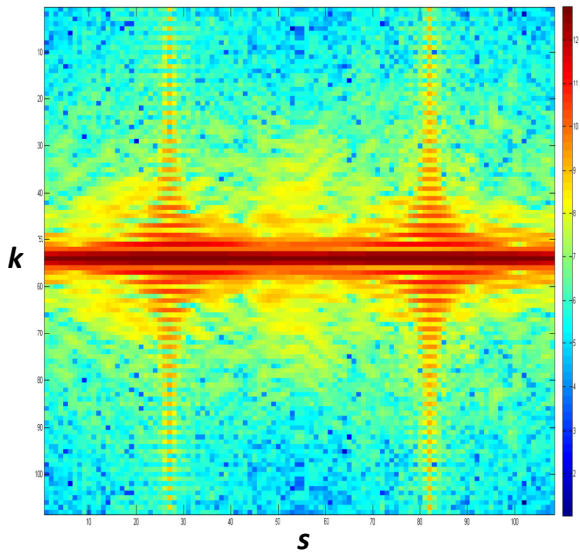

Fig 3. The *k* space picture of Fig 2. in the pseudopolar grids in log scale.

form denoise method gets less noise than the Cartesian method and the pseudopolar transform denoise method affects more critically in the wedge than the center, so if I only calculate the noise/signal of 52×52 pixels in the center between two methods. I can find the noise/signal=0.0476 and noise/signal=0.0498 of my method and the normal method, respectively. That means my method is more successful than the normal way in the center of the real space pictures. I also show the variances of each column between my method and normal method in the real space results. I can see my method is always low variances the normal way both in the maximum and minimum variances, unless the last two columns of the minimum variances, let's see Fig. 6 and 7. I are now doing *N*×*N*=156×156 pixels picture noise removal, and it can be shown in the conference this year. A 300×300 pixels picture is not hard to be done by my method. For any *N*×*N* matrix of the exact inverse pseudopolar Fourier transform, we only need to do one time and save the matrix to the hard disk as a library for anyone, it save much time that we won't do repeated work again. In the end, thank Dr. Kung-Hsuan Lin and Dr. Ting-Kuo Lee from Academia Sinica for helpful discussions.

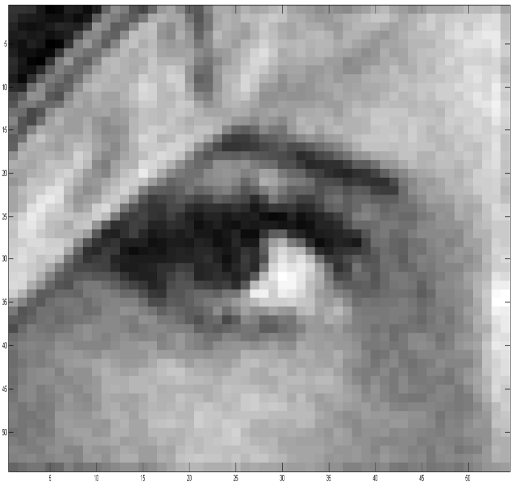

Fig 4. The picture calculated by the exact inverse pseudopolar grid Fourier transform after doing Gaussian filter in the *k* space, and the noise/signal=0.0525.

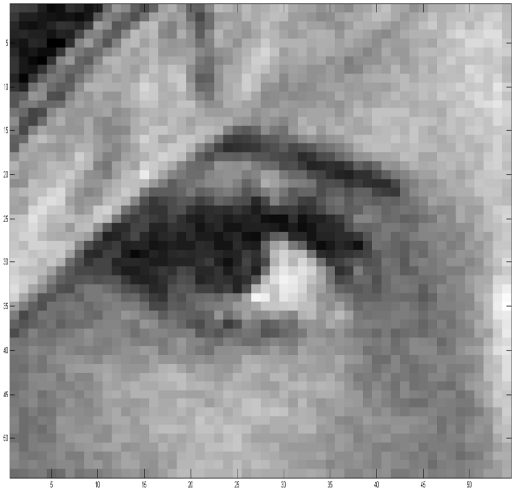

Fig 5. The picture calculated by the normal Cartesian grid inverse Fourier transform also after doing Gaussian filter in the *k* space, and the noise/signal=0.0544.

## 5. MORE DISCUSSIONS

As for the comparison, I also did the normal Cartesian grid fast Fourier transform, and also choose the same filter, Gaussian filter, to remove noise in the *k* space of a Cartesian grid. I find the noise/signal=0.0544 in Fig 5., which is worse than the noise removals in the pseudopolar Fourier space, but is the best noise/signal of the normal method with different Gaussian function parameters. We can see Fig 4. is more smooth than Fig 5., that means the pseudopolar trans-

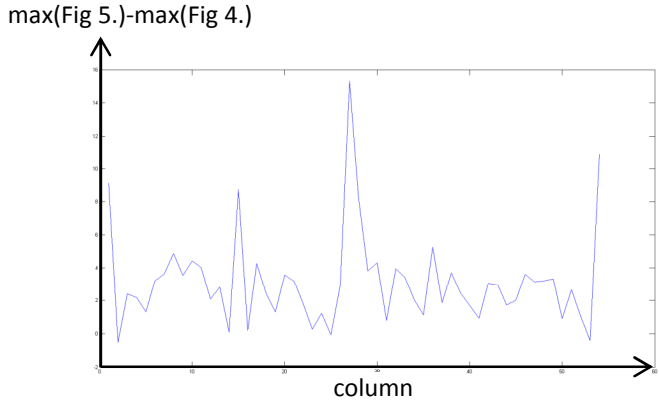

Fig 6. The max(Fig 5.)-max(Fig 4.) of each column of Fig 5. and Fig 4. in the real space.

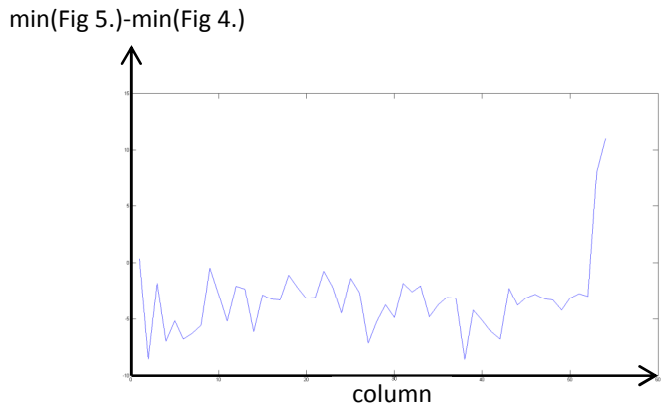

Fig 7. The min(Fig 5.)-min(Fig 4.) of each column of Fig 5. and Fig 4. in the real space.